\documentclass[conference, final]{IEEEtran}
\IEEEoverridecommandlockouts
\usepackage{amsmath,amssymb,amsfonts}
\usepackage{algorithmic}
\usepackage{graphicx}
\usepackage{textcomp}
\usepackage{xcolor}
\usepackage{bm}
\usepackage{subfigure}
\usepackage{hyperref}
\newcommand{\angstrom}{\mbox{\normalfont\AA}}

\def\BibTeX{{\rm B\kern-.05em{\sc i\kern-.025em b}\kern-.08em
    T\kern-.1667em\lower.7ex\hbox{E}\kern-.125emX}}
\begin{document}

\title{Attention-Based Generative Neural Image Compression on Solar Dynamics Observatory\\
\thanks{This research is based upon work supported by the National Aeronautics and Space Administration (NASA), via award number 80NSSC21M0322 under the title of \emph{Adaptive and Scalable Data Compression for Deep Space Data Transfer Applications using Deep Learning}.}
}

\author{
    Ali Zafari$^\dag$, Atefeh Khoshkhahtinat$^\dag$, Piyush M. Mehta$^\ddag$, Nasser M. Nasrabadi$^\dag$, Barbara J. Thompson$^\S$,\\ Daniel da Silva$^\S$, Michael S. F. Kirk$^\S$\\
    $^\dag$Dept. of Computer Science \& Electrical Engineering, West Virginia University, WV USA\\
    $^\ddag$Dept. of Mechanical \& Aerospace Engineering, West Virginia University, WV USA\\
    $^\S$NASA Goddard Space Flight Center, MD USA\\
    %Institution1 address\\
    {
    \tt\small \{\href{mailto:az00004@mix.wvu.edu}{az00004},\href{mailto:ak00043@mix.wvu.edu}{ak00043}\}@mix.wvu.edu,\{\href{mailto:piyush.mehta@mail.wvu.edu}{piyush.mehta},\href{mailto:nasser.nasrabadi@mail.wvu.edu}{nasser.nasrabadi}\}@mail.wvu.edu
    % \tt\small \{\href{mailto:barbara.j.thompson@nasa.gov}{barbara.j.thompson}, \href{mailto:daniel.e.dasilva@nasa.gov}{daniel.e.dasilva}, \href{mailto:michael.s.kirk@nasa.gov}{michael.s.kirk}\}@nasa.gov
    % }
    }\\
    {
    \tt\small \{\href{mailto:barbara.j.thompson@nasa.gov}{barbara.j.thompson},\href{mailto:daniel.e.dasilva@nasa.gov}{daniel.e.dasilva},\href{mailto:michael.s.kirk@nasa.gov}{michael.s.kirk}\}@nasa.gov
    
    }
}

% \author{\IEEEauthorblockN{Ali Zafari}
% \IEEEauthorblockA{\textit{Dept. of Computer Science \& Electrical}\\
% \textit{Engineering, West Virginia University}\\
% Morgantown, WV USA \\
% \href{mailto:az00004@mix.wvu.edu}{az00004@mix.wvu.edu}}
% \and
% \IEEEauthorblockN{Atefeh Khoshkhahtinat}
% \IEEEauthorblockA{\textit{Dept. of Computer Science \& Electrical}\\
% \textit{Engineering, West Virginia University}\\
% Morgantown, WV USA \\
% \href{mailto:ak00043@mix.wvu.edu}{ak00043@mix.wvu.edu}}
% \and
% \IEEEauthorblockN{Piyush M. Mehta}
% \IEEEauthorblockA{\textit{Dept. of Mechanical and Aerospace} \\
% \textit{Engineering, West Virginia University}\\
% Morgantown, WV USA \\
% \href{mailto:piyush.mehta@mail.wvu.edu}{piyush.mehta@mail.wvu.edu}}
% \and
% \IEEEauthorblockN{Nasser M. Nasrabadi}
% \IEEEauthorblockA{\textit{Dept. of Computer Science \& Electrical}\\
% \textit{Engineering, West Virginia University}\\
% Morgantown, WV USA \\
% \href{mailto:nasser.nasrabadi@mail.wvu.edu}{nasser.nasrabadi@mail.wvu.edu}}
% \and
% \IEEEauthorblockN{Barbara J. Thompson}
% \IEEEauthorblockA{\textit{NASA Goddard Space Flight Center}\\
% Greenbelt, MD USA \\
% \href{mailto:barbara.j.thompson@nasa.gov}{barbara.j.thompson@nasa.gov}}
% \and
% \IEEEauthorblockN{Daniel da Silva}
% \IEEEauthorblockA{\textit{NASA Goddard Space Flight Center}\\
% Greenbelt, MD USA \\
% \href{mailto:daniel.e.dasilva@nasa.gov}{daniel.e.dasilva@nasa.gov}}
% \and
% \IEEEauthorblockN{Michael S. F. Kirk}
% \IEEEauthorblockA{\centerline{\textit{NASA Goddard Space Flight Center}}\\
% Greenbelt, MD USA \\
% \href{mailto:michael.s.kirk@nasa.gov}{michael.s.kirk@nasa.gov}}
% }
\IEEEpubid{\begin{minipage}{\textwidth}\ \\[12pt]
 978-1-6654-6283-9/22/\$31.00 \copyright 2022 IEEE\\ 
 DOI 10.1109/ICMLA55696.2022.00035
\end{minipage}} 

\maketitle

\begin{abstract}
% NASA's Solar Dynamics Observatory (SDO) mission gathers 1.4 terabytes of data each day from its geosynchronous orbit in space. SDO data includes images of the Sun captured at different wavelengths, with the primary scientific goal of understanding the dynamic processes governing the Sun. To transmit this huge amount of data, SDO accomplishes this by having a dedicated ground antenna that allows a continuous downlink of data of 150 megabits per second. Unlike other solar missions, SDO uses lossless compression. However, future missions are planned that will have with much more limited data downlink budgets. To achieve high scientific performance with a smaller data transmission volume, these future missions must compress the raw data efficiently.  (BJT) (done/Ali)

%Solar Dynamics Observatory (SDO) mission of NASA gathers 1.4 Tera-bytes of data each day orbiting in space. To transmit this huge amount of data to be used in downstream scientific research, it is required to compress the raw data efficiently. This data includes images of the Sun captured at different wavelengths.
NASA's Solar Dynamics Observatory (SDO) mission gathers 1.4 terabytes of data each day from its geosynchronous orbit in space. SDO data includes images of the Sun captured at different wavelengths, with the primary scientific goal of understanding the dynamic processes governing the Sun.
Recently, end-to-end optimized artificial neural networks (ANN) have shown great potential in performing image compression. ANN-based compression schemes have outperformed conventional hand-engineered algorithms for lossy and lossless image compression. We have designed an ad-hoc ANN-based image compression scheme to reduce the amount of data needed to be stored and retrieved on space missions studying solar dynamics. In this work, we propose an attention module to make use of both local and non-local attention mechanisms in an adversarially trained neural image compression network. We have also demonstrated the superior perceptual quality of this neural image compressor. 
% We train the model on the SDO images that have not been subjected to lossy compression, in order to study the performance of lossy compression mechanisms.  (BJT) (done/Ali)
Our proposed algorithm for compressing images downloaded from the SDO spacecraft performs better in rate-distortion trade-off than the popular currently-in-use image compression codecs such as JPEG and JPEG2000. In addition we have shown that the proposed method outperforms state-of-the art lossy transform coding compression codec, i.e., BPG.

%The most important advantage of compressing images using neural networks, in comparison to conventional methods, is that we directly optimize Rate-Distortion criterion. By enforcing the target bit-rate which is defined by the capacity of the channel or the available amount of storage for archiving each image, we can optimize the parameters of the artificial neural network to reach the targeted bit rate. This job is being done by optimizing the trade-off between rate and distortion, in which the distortion is measured by a distance metric between the uncompressed image and its corresponding reconstruction. To make the distribution of reconstructions and uncompressed images closer to each other, we adopted adversarial training in the framework of generative adversarial network (GAN) as well.\\
%These neural networks could be trained on smaller crops of images to reach a targeted bit rate, and then be used to compress images at any given size to that bit rate. Even if different resolutions of the solar observations need to be accessible to the solar community for scientific purposes, we train only one neural network to compress images of different sizes at the desired quality.
\end{abstract}

\begin{IEEEkeywords}
Learned lossy image compression, solar dynamics observatory, generative adversarial network, attention
\end{IEEEkeywords}

\section{\textbf{Introduction}}
Image compression using artificial neural networks (ANN) has shown great potential to be applied on a wide variety of different areas since their first appearance \cite{toderici2015}. 
% In the past couple of years (BJT) (done/Ali)
% In just about few years 
In the past couple of years, they have outperformed most of the hand-engineered codecs such as JPEG \cite{jpeg} and JPEG2000 \cite{jpeg2000} in terms of rate-distortion (RD) performance \cite{yang2022introduction}. One major advantage of ANN-based compression algorithms is that they can be developed on any ad-hoc dataset to do better compression than general codecs \cite{yang2022introduction}.

Although it is believed that the ultimate trade-off in image compression is between the rate and distortion, recent studies have shown that there is a third role governing the visual quality of compressed images known as perception \cite{blau2019}. Generative Adversarial Networks (GANs) are known for their high-quality reconstructed images by enforcing the ANN to capture the distribution of their input image. Hence, to improve the perceptual quality of reconstructed image at the receiver, GANs have been applied to image compression networks in the literature \cite{mentzer2020}.

\begin{figure}[tp]
    \centering
    \includegraphics[width=0.9\linewidth]{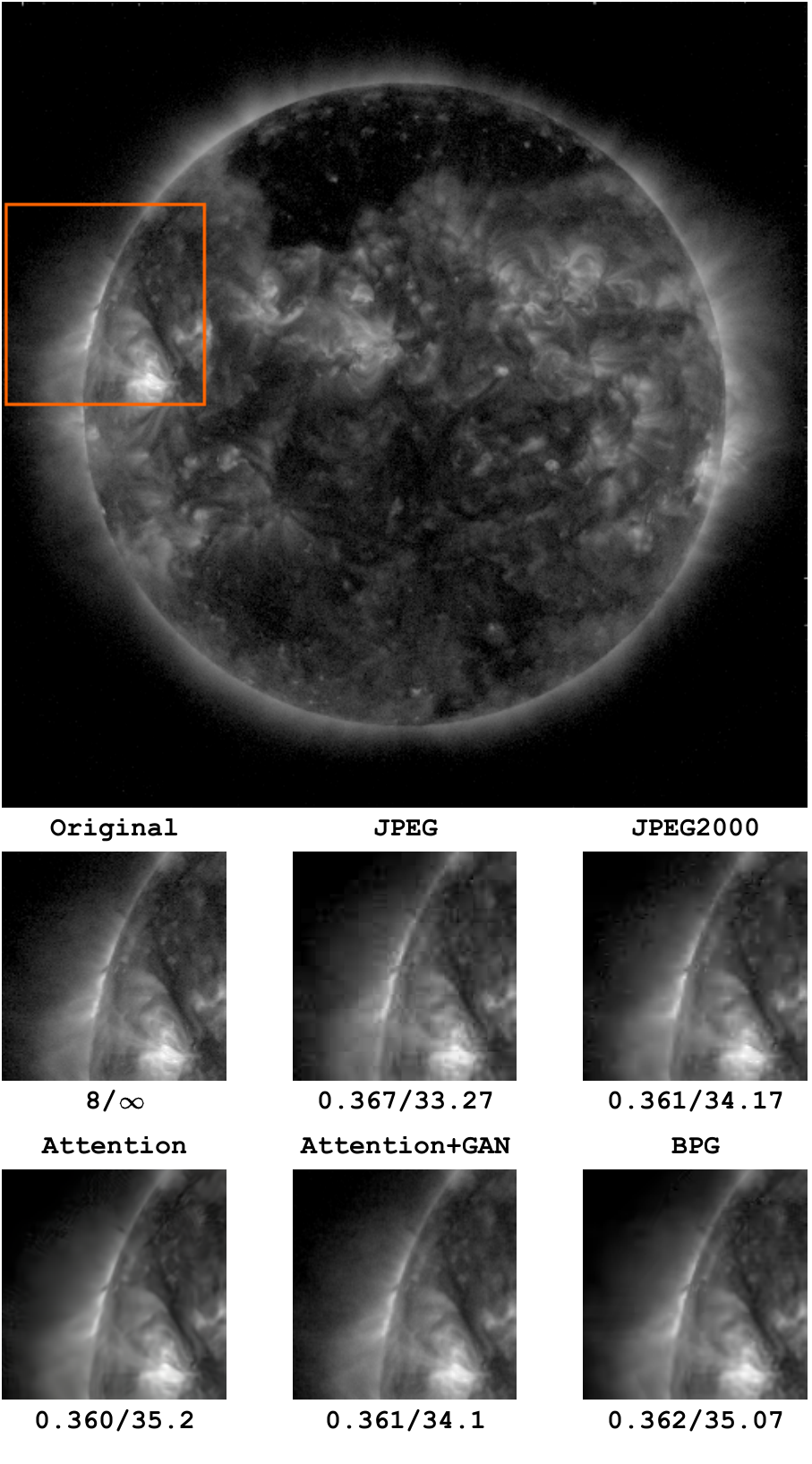}
    \caption{Visual comparison of proposed compression schemes (Attention only and GAN+Attention) to other standard codecs. Reported performance in terms of bit-rate/distortion [bpp$\downarrow$/PSNR$\uparrow$]. GAN outputs are visually closer to the original input unless their lower performance in terms of PSNR. \emph{Best viewed on screen.}}
    \label{fig:visual-comparison}
\end{figure}

Another venue of works to improve the performance of Convolutional Neural Networks (CNNs) is attention mechanism. With its unprecedented influence in natural language processing \cite{vaswani2017transformer}, attention has found its way in computer vision and object detection/classification tasks \cite{dosovitskiy2021vit,liu2021}. We have utilized both of these improvements in learned image compression networks to enhance the performance in terms of rate-distortion-perception trade-off \cite{blau2019}. As shown in Figure \ref{fig:visual-comparison}, although the attention mechanism can reach better performance compared with other compression standards, augmenting it with a GAN will lead to better perceptual quality.

\textbf{Contributions of This Work}. In this work we have investigated the application of recently successful learned image compression methods in the field of solar imaging. We have shown that these neural compression schemes could easily outperform traditional and currently-in-use image codecs. In addition, we have proposed a curated attention module to improve the RD tradeoff performance in state-of-the art neural compression architectures. We have also utilized adversarial training to encourage the decoder of our neural network to preserve the distribution of the solar images during the reconstruction process.

The remainder of the paper is organized as follows. Section \ref{sec:related-work} reviews the neural-based compression methods and the importance of compression on SDO mission. Section \ref{sec:methods} describes our proposed method. The experiments and ablation studies are discussed in section \ref{sec:experiments} with a conclusion at section \ref{sec:conclusion}.

\section{\textbf{Related Work}}\label{sec:related-work}

\subsection{\textbf{Neural Image Compression}}
Transform coding based image compression algorithms share four main steps to compress an image \cite{goyal2001transformcoding}. 
% SDO data is not RGB, it is a continuous scale at 16 bits per pixel, just counts and not colorized (BJT) (done/Ali)
% RGB here is mentioned just as an example for the unfamiliart reader, to understand quickly what's going on in this paper (Ali)
First, encoding the images from their input space (e.g., RGB) to an uncorrelated space. Second, quantizing to discard less significant information from the data in its uncorrelated domain. At the third step, an entropy coding will be utilized to losslessly encode the quantized samples into a stream of ones and zeros. This bitstream will be the compressed image. Final step occurs at the receiving end (or at reconstructing step), which is responsible to decode the quantized values to the original space of the input image. The first and most widely used architecture to mimic this scenario in deep neural networks, is the convolutional autoencoder, which has shown its superiority in the literature \cite{balle2017endtoend}.
Both the encoding and decoding part of the traditional transform coding, could be imitated by an autoencoder \cite{balle2021}.

End-to-end optimizing of the neural networks are capable of handling various tasks \cite{DEHKORDI2022109091, ebrahimi2022, akyash2021, KASHIANI201917} if the learning objective chosen to be differentiable. In an end-to-end optimisation of an autoencoder, problems arise when we want to do quantization on its bottleneck. It is worth mentioning that quantization is the essential part of compression. Merely doing dimensionality reduction cannot necessarily result in discarding the redundant information, which is necessary to attain high compression ratios \cite{theis2017}.

ANNs are optimized using gradient descent algorithms which update the parameters of the network by back-propagating the gradients of the loss function. Thus, all the operations performed in it must be differentiable. As a result, we need to approximate hard discrete quantization with a soft continuous operation. To do so, several approaches have been proposed in the literature. \cite{toderici2015} used recurrent neural networks to directly binarize the latent code stochastically, while \cite{theis2017} used an approach similar to straight through estimator \cite{bengio2013} by back-propagating the gradients of identity function and rounding to the nearest integer in the forward pass. By this continuous approximation, the network parameters can be successfully learned with backpropagating the gradient of the loss function.
%Soft-to-hard quantization was another solution to make the gradients at the first steps of 
%training more informational than in the end ???  (BJT) (done/Ali)
%training more informational than in the end \cite{agustsson2017}. 
The most widely used approach is proposed by \cite{balle2016}, inherited from \cite{gray1998}, they showed that adding independent and identically distributed uniform noise in the range of scalar quantization can be interpreted equivalently as doing scalar quantization on the bottleneck. By doing so, we can optimize the differential entropy of the continuous approximation as a variational upper bound \cite{theis2016} to reduce the entropy of the bottleneck. Low entropy messages are compressed more efficiently into bitstreams \cite{cover1999infotheory, aghdaie2022morph}.

In classical image compression schemes, to get the best out of the quantization process, the first step was to apply an invertible linear transform and translate the image into  decorrelated coefficients using Discrete Cosine Transform (DCT). By doing so, scalar quantization could reach a reasonable performance close to vector quantization \cite{goyal2001transformcoding}. The application of vector quantization in ANN-based compression has been investigated by  \cite{agustsson2017}, with the cost of complicated training procedure. On the other hand, it has been shown \cite{balle2017endtoend, balle2021} that a joint-optimized learned nonlinear transform, i.e., neural network, followed by scalar quantization can ideally approximate a parametric form of vector quantization.

Replacing the actual quantization with uniform noise approximation in the bottleneck of a vanilla autoencoder during training of the network
\cite{balle2018a}, will transform it to a Variational Aueoencoder (VAE) \cite{kingma2014}.
%It is worth noting that optimizing an autoencoder for image compression by the uniform noise approximate for the quantization \cite{balle2018a}, will lead to optimizing a variational autoencoder (VAE).
The only difference is on the chosen prior. In autoencoder-based image compression, the Gaussian prior of the VAE is replaced with unit uniform distribution centered on integer numbers.

\subsection{\textbf{Solar Dynamics Observatory (SDO) Mission}}
\subsubsection{\textbf{Image Compression on SDO Data}}
Advances in sensor technology and an increasing desire for a deeper understanding of the space environment (Sun to Earth and beyond) have led to an explosion of data volume in recent years (unprecedented spatial and/or temporal resolution as well as multi-spectral data). As a result, it requires new innovative data compression algorithms. The SDO mission transmits 1.4 TB of data (most of it images of the sun at different wavelengths) each day to the ground station \cite{sdoguide}, which shows the importance of compression on transmitting and more importantly on archiving this huge amount of data \cite{chamberlin2012solar}. 
% SDO only does lossless compression, which is why we are training on SDO data.  To get comparable performance to SDO in the future, we need good lossy compression.  (BJT) (done/Ali)
In \cite{fischer2017jpeg2000eve} authors pointed out the essential need to do lossy image compression on petabyte size data gathered from Solar missions. They studied usage of JPEG2000 which is a transform coding compression scheme based on discrete wavelength transform on SDO data.

\subsubsection{\textbf{Imagery Instruments on SDO Spacecraft}}
% SDO data are captured using three instruments onboard that gather data from the Sun 24 hours a day. The \emph{Helioseismic and Magnetic Imager} (HMI) was created to investigate oscillations and the magnetic field at the solar surface, or photosphere \cite{schou2012hmi}. The \emph{Atmospheric Imaging Assembly} (AIA) on SDO takes full-sun images (1.3 solar diameters) of the solar corona at a spatial resolution of near 1 arcsec, with an image cadence of 12 seconds for multiple wavelengths \cite{lemen2011aia}. To better understand variations on the timeframes that affect Earth's climate and near-Earth space, the Extreme ultraviolet Variability Experiment (EVE) analyzes the solar Extreme UltraViolet (EUV) irradiance with high spectral precision \cite{woods2010eve}.  (BJT) (done/Ali)
SDO data are captured using three instruments onboard that gather data from the Sun 24 hours a day. The \emph{Helioseismic and Magnetic Imager} (HMI) was created to investigate oscillations and the magnetic field at the solar surface, or photosphere \cite{schou2012hmi}. The \emph{Atmospheric Imaging Assembly} (AIA) on SDO takes full-sun images (1.3 solar diameters) of the solar corona at a spatial resolution of near 1 arcsec, with an image cadence of 12 seconds for multiple wavelengths \cite{lemen2011aia}. To better understand variations on the timeframes that affect Earth's climate and near-Earth space, the \emph{Extreme ultraviolet Variability Experiment} (EVE) analyzes the solar Extreme UltraViolet (EUV) irradiance with high spectral precision \cite{woods2010eve}. 

\subsubsection{\textbf{Machine/Deep Learning on SDO Data}}
Recently \cite{galvez2019} has gathered a portion of SDO raw data and cleaned it as machine-learning ready dataset to be used in developing new learning-based methods on SDO mission data. From here now on, we call this dataset \emph{SDOML}. Based on this dataset,  \cite{salvatelli2022} used a U-Net in a GAN network to translate AIA multi-spectral (94, 171, 193 \angstrom) images to a specific wavelength (211 \angstrom). As another machine learning work on SDOML dataset, \cite{santos2020} proposed deep neural networks as a means to auto-calibrate the instrument degradation on SDO imagery instruments. A conditional GAN is used in \cite{dash2021super} to translate downloaded HMI images from SDO to AIA images. More details on the SDOML dataset will be presented in section \ref{sec:experiments:dataset}.

\section{\textbf{Methods}}\label{sec:methods}
\begin{figure*}[tp]
    \centering
    \includegraphics[width=0.9\linewidth]{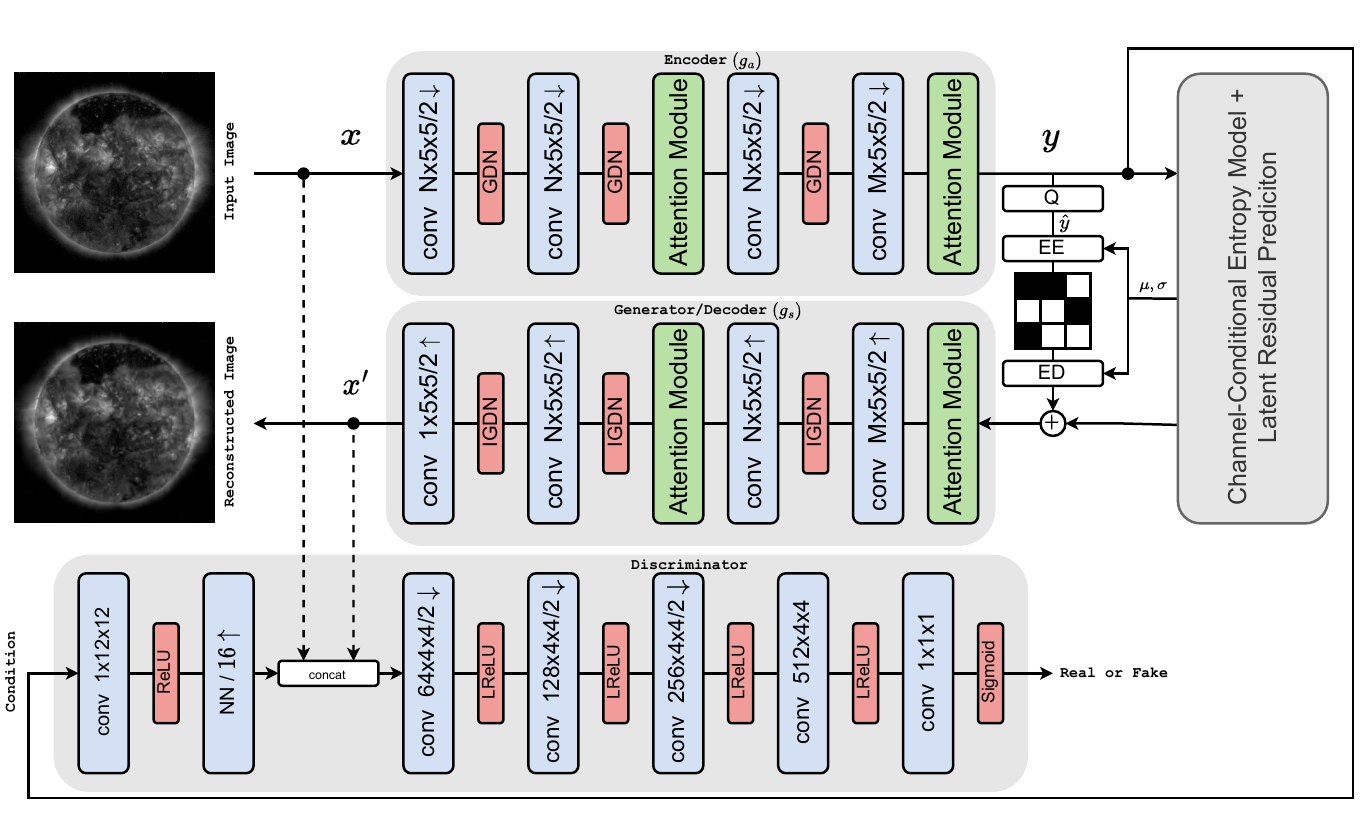}
    \caption{Network architecture. Input image is down-scaled by a factor of 16 to get the latent code and up-sampled in reverse to get the reconstructed image. A conditional discriminator encourages the generator (decoder) for better perceptual quality. Number of channels in encoder and decoder are set by $N=192$ and $M=320$. Q performs the scalar quantization. EE and ED indicate entropy encoder and decoder, respectively. $\mu$ and $\sigma$ are predicted parameters of the latent code probability distribution, defined by entropy estimation model. GDN and IGDN correspond to Generalized Divisive Normalization nonlinearity and its inverse, discussed in section \ref{sec:experiments:implementation}.}
    \label{fig:network-arch}
\end{figure*}

\subsection{\textbf{Generative Image Compression}}
Autoencoder based learned image compression networks, like the one we have proposed in Figure \ref{fig:network-arch}, generally consist of two major parts. First the encoder/decoder network, and second the bottleneck entropy estimation network. The latter is discussed in depth in section \ref{section:entropy-model}. According to Figure \ref{fig:network-arch}, the network input ($\bm{x}$) and output ($\bm{x'}$) relations can be summarized as follows

\begin{equation}
 \begin{aligned}
    &\bm{x'}=g_s(\bm{\hat{y}};\bm{\theta_g}),\\
    &\bm{\hat{y}}=\lfloor g_a(\bm{x};\bm{\phi_g})\rceil,\\
    &\bm{\hat{z}}=\lfloor h_a(\bm{y};\bm{\phi_h})\rceil,
\end{aligned}
\end{equation}
in which, $\lfloor\cdot\rceil$ denotes quantizing the real valued input to the nearest integer number. $\bm{\hat{y}}$ is the quantized latent variable and $\bm{\hat{z}}$ is its quantized hyper-prior. Encoder and decoder nonlinear transforms are represented by $g_a$ and $g_s$ with their learned parameters, $\bm{\phi_g}$ and $\bm{\theta_g}$, respectively. The subscripts $a$ and $s$ refer to \emph{analysis} and \emph{synthesis} as they are common words in the area of transform coding based compression. $h_a$ is the analysis transform to get the hyper-priors of the entropy estimation model, leaned by its parameters $\bm{\phi_h}$.

\subsubsection{\textbf{Learning Objective}} \label{sec:compression:objective}
Any learned image compression network tries to tackle with rate-distortion trade-off, governed by a Lagrangian coefficient $\lambda$ which can be described as
\begin{equation}
    R+\lambda D
    \label{eq:rate-distortion},
\end{equation}
where $R$ and $D$ correspond to the estimated entropy of the latent code and reconstruction distortion, respectively.
Estimated entropy of the quantized bottleneck represents the rate term which is desired to be minimized during the training of the neural network. The probability distribution of the latent code is variationally approximated by hyper-prior $\bm{z}$. Then the quantized $\bm{\hat{z}}$ is transmitted alongside the compressed image as a side-information. Therefore, the entropy of both should be optimized as defined below
\begin{equation}
    R=\mathbb{E}_{x\sim p_X}[-\log_2P_{\bm{\hat{y}}|\bm{\hat{z}}}(\bm{\hat{y}}|\bm{\hat{z}};\bm{\theta_h})-\log_2P_{\bm{\hat{z}}}(\bm{\hat{z}};\bm{\psi})],
\end{equation}
where $\bm{\theta_h}$ and $\bm{\psi}$ are parameters of learned entropy model on latent code ($\bm{\hat{y}}$) and hyper-prior ($\bm{\hat{z}}$), respectively.

In Eq. \ref{eq:rate-distortion}, $D$ accounts for the distortion between input and output image of the network which can be measured by any desired metric. The prevalently used criterion to measure distortion between input and output is the Mean Squared Error (MSE), which is heavily criticized because of reconstructing blurry images. Efforts have been made to propose metrics which can adhere perceptually to human visual system, e.g. Multi Scale Structural SiMilarity Index (MS-SSIM) \cite{wang2009}. Even these metrics have shown weaknesses when intensely scrutinized \cite{nilsson2020}.
% I'm not sure what "reconstructing blurry images" is, but my primary concern about MSE is that it prioritizes pixel-to-pixel brightness and brighter regions, which cover a small fractional area of the sun.  Regions with less emission do not contribute enough to MSE but most of the solar atmosphere is lower emission.  (BJT) (done/Ali)
%SSIM \cite{wang2004} and MS-SSIM \cite{wang2003}.

Recently, perceptual-aware metrics based on features generated by pre-trained neural networks have been proposed. Learned Perceptural Image Patch Similarity (LPIPS) introduced by \cite{zhang2018lpips} uses trained AlexNet/VGGNet features to compare patches of an image with a corresponding reference. In training our neural compressor we will fortify its reconstruction loss by exploiting this perceptual metric.

To make the reconstruction closer to the input image, we also consider adversarial training of our decoder network. Generative Adversarial Networks (GANs) \cite{goodfellow2014} consisting of a generator and a discriminator sub-network, are able to follow the distribution of data at reconstruction instead of just trying to find the nearest pixel values in order to decrease the distortion. In our network the decoder plays the role of the generator. Then the discriminator forces the decoder output to preserve the distribution of the input image at the reconstructed image. The proposed objective to be optimized is a combination of distortion and perception as follows
\begin{equation}
    \begin{split}
    D=\mathbb{E}_{\bm{x}\sim p_X}[&\lambda_{recon}MSE(\bm{x},\bm{x'})\\+&\lambda_{perc}LPIPS(\bm{x},\bm{x'})\\-&\lambda_{adv}\log D(\bm{x'},\bm{y})].
    \end{split}
\end{equation}
To make the adversarial training feasible we need the discriminator to judge whether its input sample came from the true distribution of data or is a fake generated one. The discriminator will need to be optimized by a separate auxiliary loss, given as
\begin{equation}
\begin{split}
    L_{disc.}=&\mathbb{E}_{\bm{x}\sim p_X}[-\log(D(\bm{x},\bm{y})]\\+&\mathbb{E}_{\bm{x}\sim p_X}[-\log(1-D(\bm{x'},\bm{y}))].
\end{split}
\end{equation}
It has been shown analytically that the distortion is in direct trade-off with perception \cite{blau2018}. GANs are the solutions to find a better perception quality by losing an acceptable amount of distortion. In \cite{blau2019} a third term in this tradeoff was introduced as the rate in the lossy compression scheme. More detailed experiments have been adopted in \cite{mentzer2020} to prove this idea in practice. Therefore, it would be an expected behavior to have a lower Peak Signal to Noise Ratio (PSNR) value on a decoder trained adversarially in contrast to a decoder trained merely on distortion metrics.

%\subsubsection{\textbf{Continuous Approximation of Quantization}} \label{sec:methods:compression:quantization}

\subsubsection{\textbf{Entropy Estimator Model}} \label{section:entropy-model}
    
Performance of any learned image compression scheme heavily depends on how well it can estimate the true entropy of the bottleneck. So the objective will be to minimize the cross entropy between the two. To make the entropy estimation possible, several probability estimation methods have been proposed in the literature, including empirical histogram density estimation \cite{agustsson2017,theis2017}, piecewise linear models \cite{balle2017endtoend}, conditioning on a latent variable (hyper-prior) \cite{balle2018a} and context modelling based on autoregressive models. \cite{minnen2018}.

%$R$ theoretically can be measured by Shannon entropy but it will be approximated by an upper-bound total information content, approximated by sum of self-information for every elemnt of the entropy-constrained bottleneck
%\begin{equation}
%    I(\Tilde{y})\approx\sum\log_2 q(\Tilde{y}_i)
%\end{equation}
%where $\Tilde{y}$ is continuous approximation of the discrete valued quantized latent codes, $\hat{y}$ from the output of the encoder.
\begin{figure*}[tp]
    \centering
    \subfigure[Attention module with skip connection. RB denote residual block.]{\includegraphics[width=0.45\textwidth]{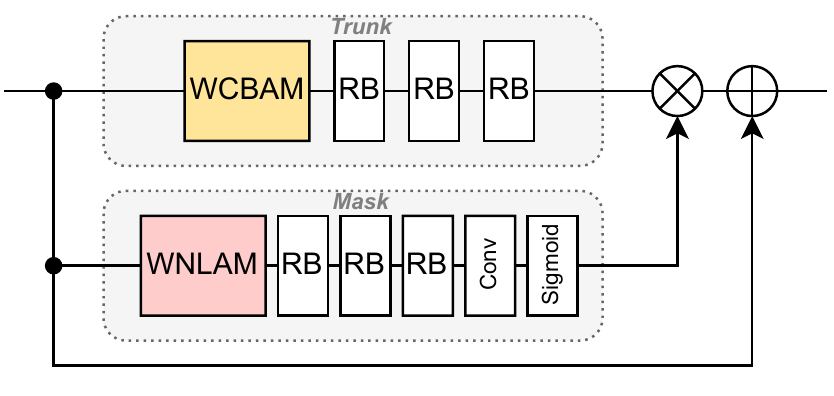} \label{fig:residual-attention}}
    \hfill
    \subfigure[Window-based non-local attention module (WNLAM).]{\includegraphics[width=0.35\textwidth]{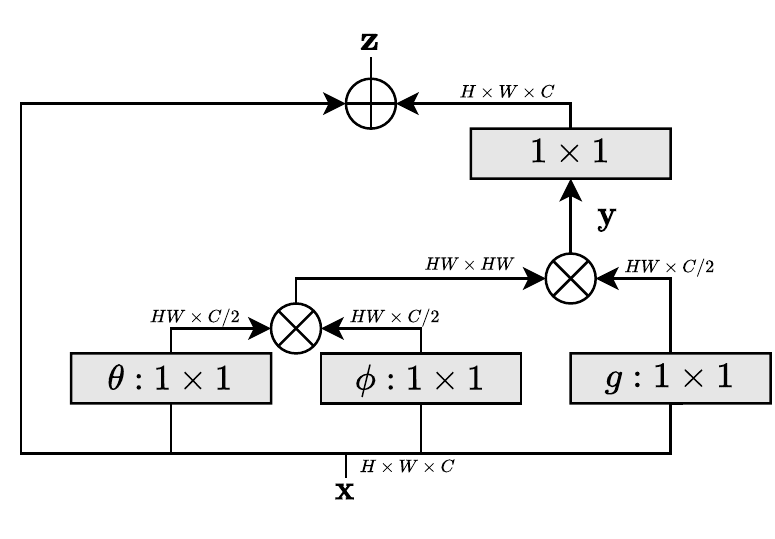}\label{fig:wnlam}}
    \vfill
    \subfigure[Window-based convolutional block attention module (WCBAM). A feature map of C channels with spatial dimensions of H$\times$W. $w$ is the window size to calculate channel attention on.]{\includegraphics[width=0.8\textwidth]{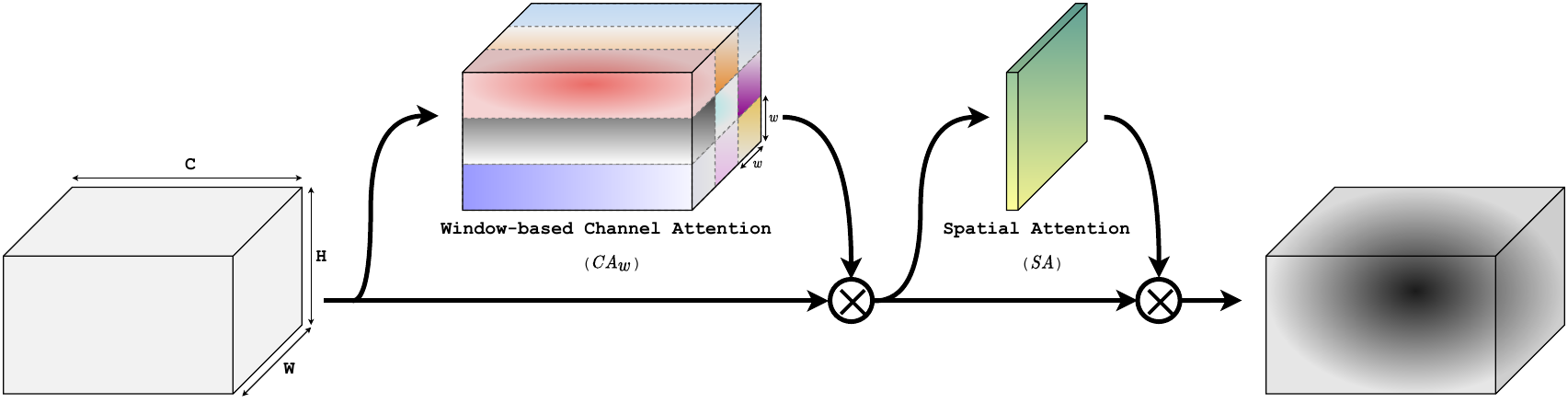} \label{fig:wcbam}}
    \caption{Attention module architecture.}
\end{figure*}

From a high-level overview, entropy estimation models can be divided into two main categories, Forward Adaptation (FA) and Backward Adaptation (BA) models. The former suffers from low capacity to capture all dependencies in the probability distribution of the latent code and the latter's disadvantage is that decoding process cannot be parallelized. Learned FA models \cite{balle2018a, balle2021} will only use the information provided during the encoding of the image, while BA methods based on autoregressive models \cite{minnen2018} need information from the decoded message as well. To take advantage of both of these models \cite{minnen2020} we define the conditional probability of the latent code as
\begin{equation}
P_{\bm{\hat{y}}|\bm{\hat{z}}}(\bm{\hat{y}}|\bm{\hat{z}})=\prod_i P(\bm{\hat{y}_i}|\bm{\hat{y}_{j<i}},\bm{\hat{z}};\bm{\theta_h}).
\end{equation}
Conditioning on quantized hyper-prior, i.e., $\bm{\hat{z}}$, as a side-information is an example of FA and conditioning on all previously decoded elements of the latent space, i.e., $\bm{\hat{y}_{j<i}}$, is an example of BA.
BA performance have been improved in \cite{minnen2020} by letting the conditioning exist only between slices of channels in the bottleneck. In contrast to spatial autoregressive modeling in \cite{minnen2018}, \cite{minnen2020} only considers the conditioning of the probabilities on the channels and it showed that by doing this the decoding process could be reasonably parallelized. We have used the same approach in \cite{minnen2020} to estimate the entropy and minimize it during the training.

\subsection{\textbf{Attention Assisted Image Compression}}

\begin{figure*}[tp]
    \centering
    \subfigure{\includegraphics[width=0.45\textwidth]{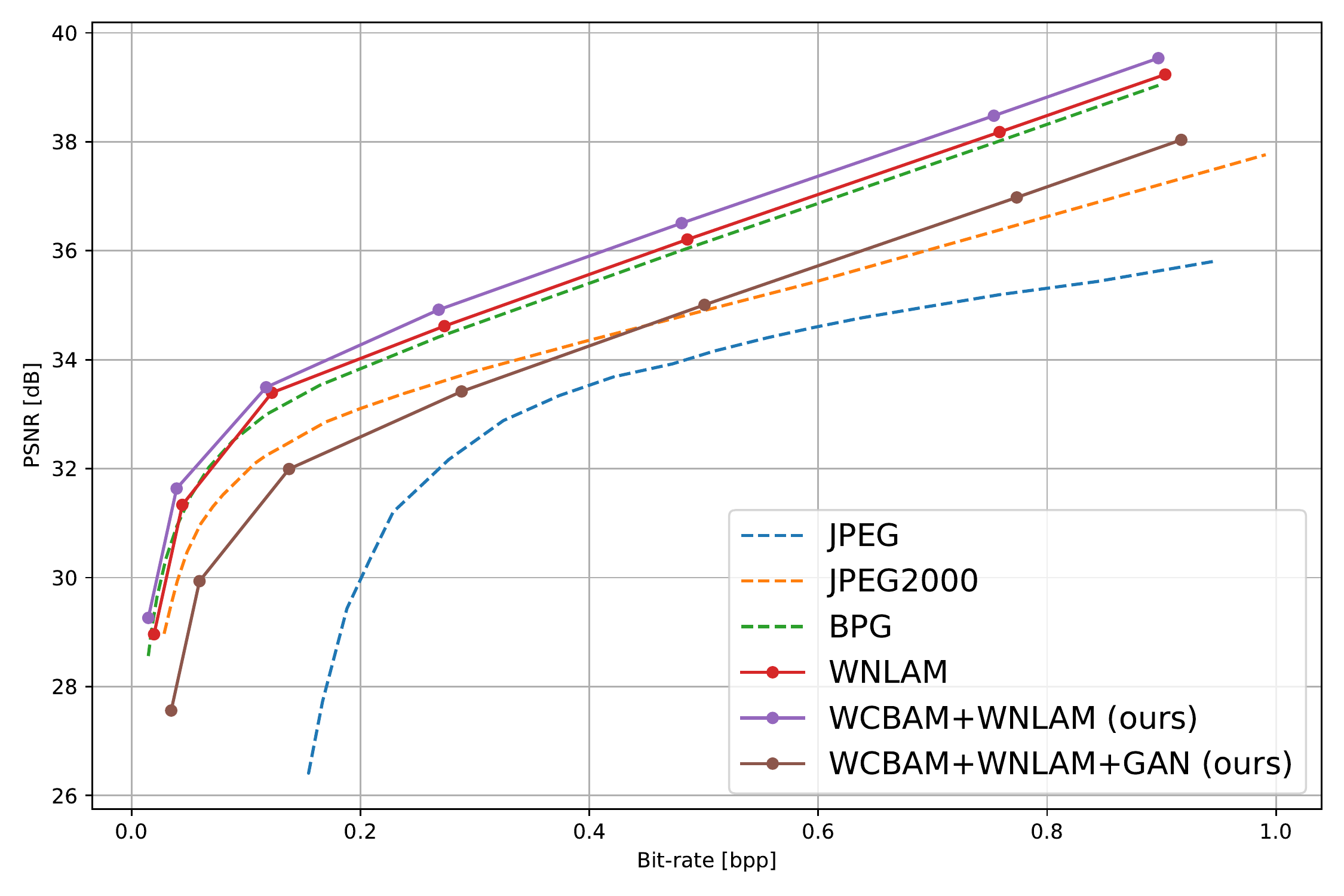}}
    \hfill
    \subfigure{\includegraphics[width=0.45\textwidth]{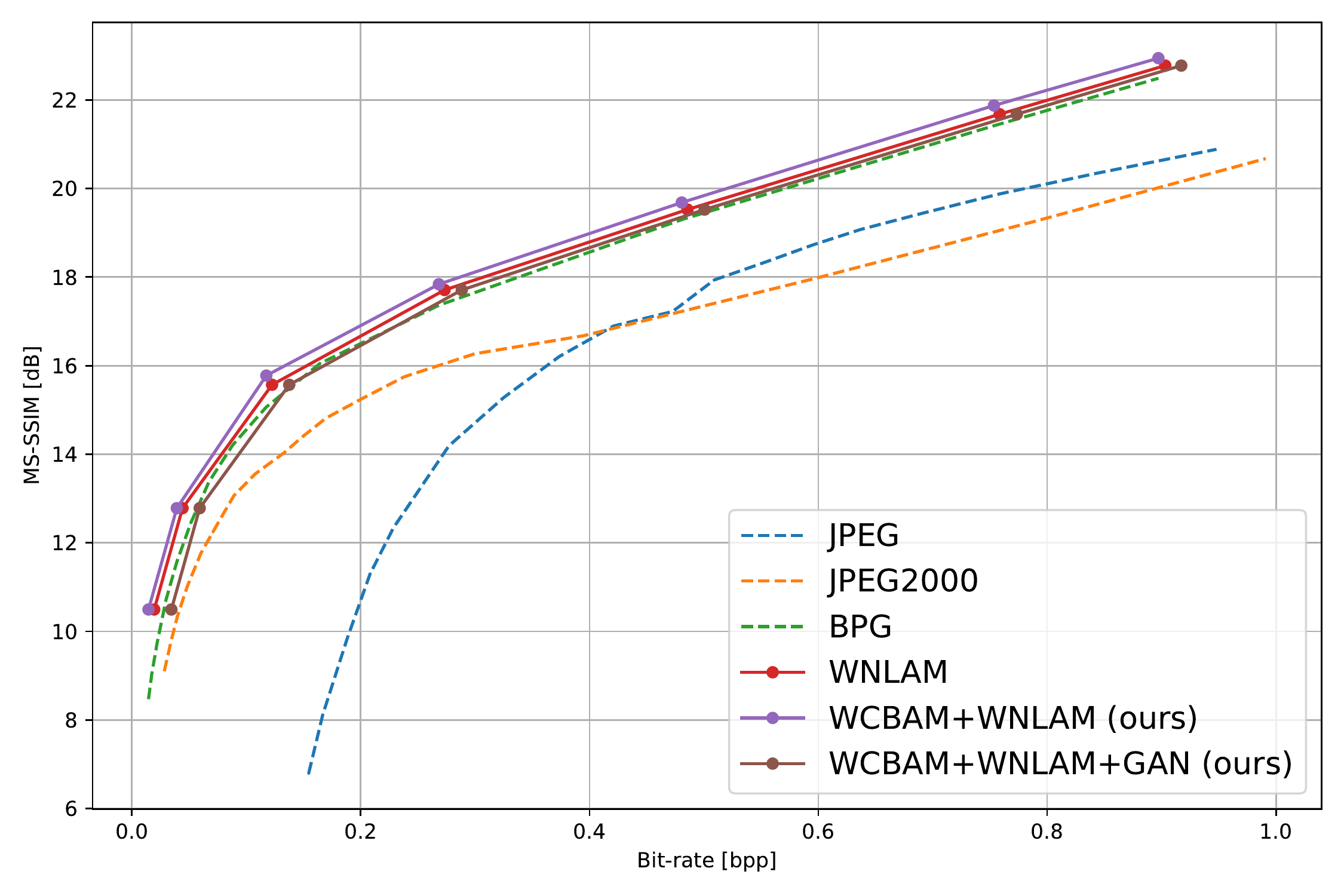}}
    \caption{Rate distortion curves aggregated over test set described on section \ref{sec:experiments:dataset}. On the left PSNR is calculated from MSE by $10\log_{10}\frac{255^2}{MSE}$. On the right MS-SSIM is reported in logarithmic scale by $-10\log(1-m)$ to show the differences better, in which $m$ is the MS-SSIM in the range of zero to one.}
    \label{fig:rd-cvrves}
\end{figure*}

When it comes to computer vision, deep convolutional neural networks are the de facto standard despite their poor performance on capturing long range dependencies. CNNs will have problems if it is required to simultaneously capture a few characteristics from non-neighboring pixels. It has been determined that the local nature of kernel sliding on just a few pixels of the input image is the primary cause of this degradation \cite{ramachandran2019stand}.

Efforts have been made to help CNNs capture more robust representation of the input image. One naive solution is to make the network deeper but other problems will arise on training such networks which has led to the introduction of deep residual networks (ResNet) \cite{he2016resnet}. Although increasing the parameters of network will generally lead to richer representation and better performance, it will make training of such networks harder.
Attention mechanisms have been proposed to address this issue of CNNs without making the network deeper. In \cite{wang2017residualattn} authors have proposed a single module to be included in between sequential convolutional layers, consisting of two branches, \emph{trunk}, to process local features and \emph{mask} to decide which of the local features in the trunk are more important to be passed to the next convolutional layer, as in Figure \ref{fig:residual-attention}.

In contrast to local attention, \cite{wang2018} first discussed how non-local attention can be viewed as a special case of non-local algorithm which was traditionally used as a method to denoise images \cite{buades2005}. The idea was to find similar pixels/patches in the image/feature map and replace it with a weighted sum over all the others, with higher weights for more similar ones. It can be inferred from  \cite{wang2018} that Vision Transformers (ViT) \cite{dosovitskiy2021vit} are all special cases of non-local attention mechanism.

Non-local attention block (Figure \ref{fig:wnlam}) helps the \emph{mask} branch to efficiently learn the most informative parts of features (in the \emph{trunk}) for the task in hand \cite{zhang2018residualnonlocal}. The authors in \cite{zhang2018residualnonlocal} also added a skip connection to help the output feature maps be richer. This skip connection prevents vanishing gradients as well. Another recently proposed simple tweak to incorporate attention in CNNs has been introduced in \cite{woo2018cbam}. It is an enhanced version of Squeeze-and-Excitation network \cite{hu2018squeeze}, to apply attention both on spatial and channels feature maps separately. This way of applying attention is simpler and computationally more efficient. 

Discussed attention mechanisms have been employed in deep learned neural compression networks as well. \cite{zhou2019attention} applied residual attention, then \cite{chen2021} improved their work by adding a non-local attention to the mask of the residual attention. To improve further, \cite{zou2022} applied non-local attention limited to small windows of the feature maps. This window-based attention attained better results in the compression area.
Here we propose to use two kinds of attention mechanisms in a window-based manner.

\subsubsection{\textbf{Window-based Non-Local Attention Module (WNLAM)}}

Non-local attention block as shown in Figure \ref{fig:wnlam} is composed of a weighted sum (weights calculated as a softmax) over linear transformed version of $\mathbf{x}$, i.e., $g(\mathbf{x})$.
\begin{equation}
    \mathbf{y}_i = \frac{1}{\sum_{\forall j}e^{\theta(\mathbf{x}_i)^T\phi(\mathbf{x}_j)}}\sum_{\forall k}e^{\theta(\mathbf{x}_i)^T\phi(\mathbf{x}_k)}g(\mathbf{x}_k),
    \label{eq:non-local}
\end{equation}
where $g(.)$ is a linear transformation ($W_g$) implemented by a $1\times1$ convolution layer defined as $g(\mathbf{x_k})=W_g\mathbf{x_k}$. The weights of the sum in Eq. \ref{eq:non-local} are calculated by the measure of similarity in embedding space of the input, i.e., $\theta(\mathbf{x}_i)=W_\theta\mathbf{x_i}$ and $\phi(\mathbf{x}_k)=W_\phi\mathbf{x_k}$.

As the final operation in non-local attention, $\mathbf{z}_i$ is calculated by a linear transformation  ($W_z$) added to the original $\mathbf{x}_i$ as follows
\begin{equation}
    \mathbf{z}_i = W_z\mathbf{y}_i+\mathbf{x}_i.
\end{equation}
In image compression, restoring edges and high-frequency content is more important than representing the global features in the latent representation. Consequently, naive non-local attention mechanism performs worse than local attentions which are able to capture local redundancies and preserve details on the reconstructed image \cite{zou2022}.

\subsubsection{\textbf{Window-based Convolutional Block Attention Module (WCBAM)}}
A simple to implement kind of attention in CNNs is the convolutional block attention module (CBAM) which has shown great benefit in classification tasks \cite{woo2018cbam}. It carries out two attention mechanisms. First, the channel attention ($CA$) guides the network to only consider channels with higher importance for the desired task. Second, the spatial attention ($SA$) dictates the network where to pay attention more. Here we propose to utilize this attention module in a window-based manner. Instead of globally considering the whole spatial dimensions of each channel, we focus only on a cropped window size of $w$, as shown in Figure \ref{fig:wcbam}.

Applying WCBAM mechanism on the input features $X$ can be summarized as
\begin{equation}
 \begin{aligned}
    X_{CA} &= CA_w \odot X,\\
    X_{CA,SA} &= SA \odot X_{CA},
\end{aligned}  
\end{equation}
where $CA_w$ reweighs the channels over each window. Then $SA$ is multiplied on each refined channel to highlight the important spatial content.

Window based channel attention is calculated by passing the average and max pool through a shared fully connected network (F), as in Eq. \ref{eq:CAw}
\begin{equation}
    CA_w = sigmoid(F(Avg(X_w))+F(Max(X_w))).
    \label{eq:CAw}
\end{equation}
Afterwards, the spatial attention weights ($SA$) will be derived by concatenating average and max pool passed through a convolutional layer as
\begin{equation}
    SA = sigmoid(Conv([Avg(X_{CA}),Max(X_{CA})])).
\end{equation}
WCBAM  helps the network to capture global dependencies which may be needed during transforming the image from pixel space to feature space, specially those which the window-based non-local attention are incapable of.

\subsubsection*{\textbf{Transformers as Attention Modules}}The superiority of models based on Transformers which are a special kind of non-local attention mechanism, has been recently proven \cite{dosovitskiy2021vit, liu2021}. Although Transformers have shown great benefit in image classification or object detection tasks, their naive application in image compression networks has failed \cite{zou2022}. The whole purpose of transformers is to capture long-range dependencies in an image, while the ultimate goal in image compression is to capture dependencies in order to summarize the dependencies efficiently in the latent code.

\section{\textbf{Experiments}}\label{sec:experiments}

\subsection{\textbf{Dataset}} \label{sec:experiments:dataset}
SDOML includes images of the sun at wavelengths 94, 131, 171, 193, 211, 304, 335, 1600, 1700 \angstrom\/ at a cadence of 6 minutes. We downsampled the images to a cadence of 1 hour to avoid the dependency between training samples.
% In addition, to prevent biases of the images with respect to solar cycles, we followed the same approach proposed by \cite{salvatelli2019} to divide the dataset based on the month they are taken. Images of January to August are chosen for training and September to December are reserved for testing. The results reported in this section are all based on this portion of dataset.
In addition, to prevent biases of the images with respect to solar variations at different stages of the solar cycle, we followed the same approach proposed by \cite{salvatelli2019} to divide the dataset based on the month they are taken. Images of January to August are chosen for training and September to December are reserved for testing. The results reported in this section are all based on this portion of dataset.
% A solar cycle is approximately 11 years, so it should say "to prevent biases of the images with respect to solar variations at different stages of the solar cycle..."  (BJT) (done/Ali)
\subsection{\textbf{Implementation Details}} \label{sec:experiments:implementation}
As nonlinearity in our neural network, we have utilized computationally efficient \cite{johnston2019gdnfast} version of Generalized Divisive Normalization (GDN) \cite{balle2016gdn}. As a result of GDN's local normalization, statistical dependencies are reduced in the feature maps. By exploiting GDN instead of more conventional nonlinearities like ReLU, the feature maps will be decorrelated. Ideally, scalar quantization of a set of decorrelated features will present compression performance close to parametric vector quantization \cite{balle2021}. During the evaluation phase, entropy coding of the latent integer values was realized by asymmetric numeral systems \cite{duda2013asymmetric}. 

Seven models have been trained with $\lambda\in\displaystyle\{0.0015, 0.0035,\\ 0.0070, 0.0125, 0.0250, 0.0410, 0.0550\}$ governing the rate-distortion trade off as in Eq. \ref{eq:rate-distortion} for 100 epochs to train each model.
We have used Adam \cite{kingma15adam} optimizer on batches of size 16 consisting of randomly cropped $256\times256$ patches out of the original images of $512\times512$. Initial value of the learning rate is set to $10^{-4}$ and annealed during the training to $1.2\times10^{-6}$.

\begin{figure}[tp]
    \centering
    \includegraphics[width=0.8\linewidth]{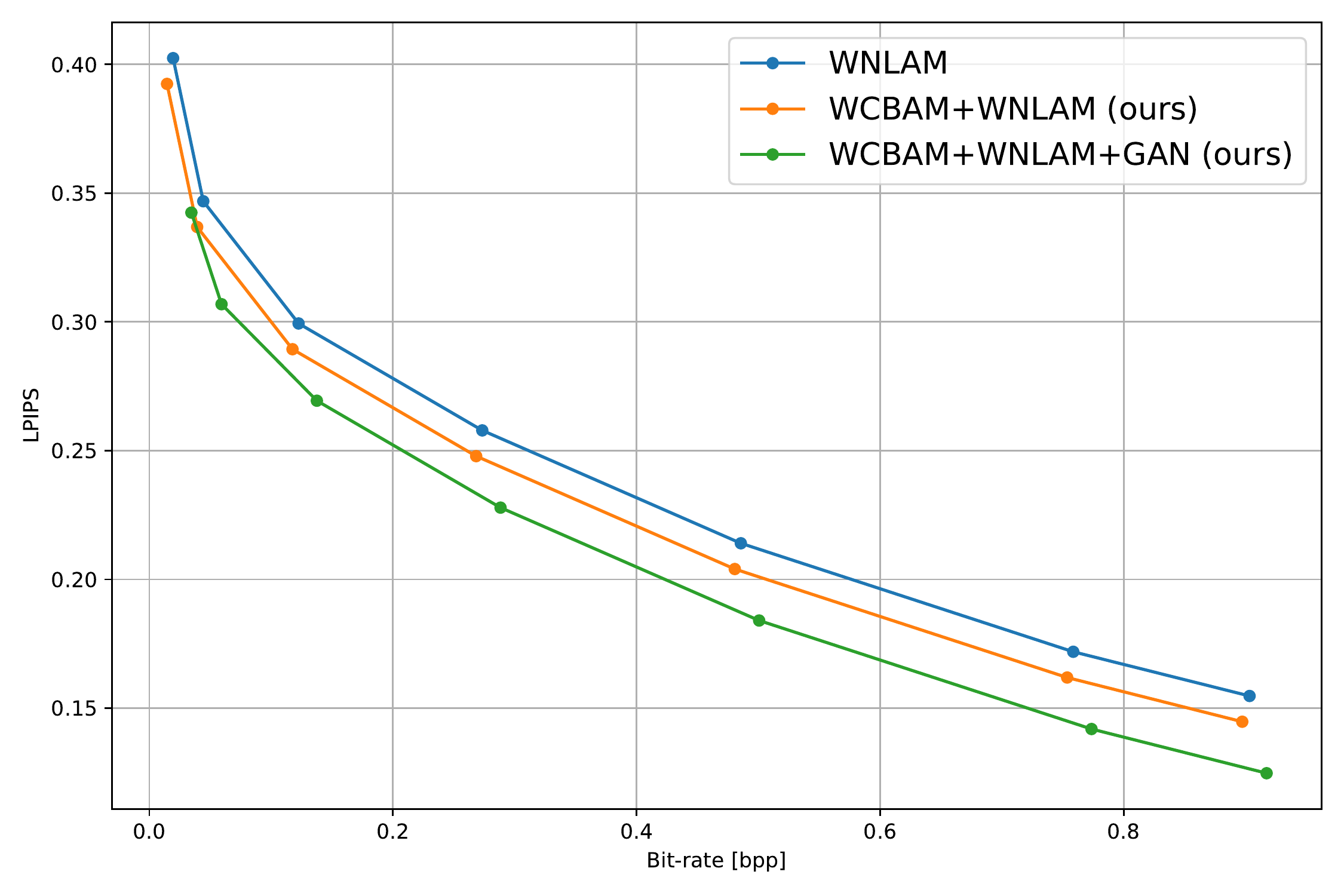}
    \caption{Rate distortion curve. Distortion is measured by LPIPS metric (lower is better) described in section \ref{sec:compression:objective}. As it can be seen, GAN performance in generating high quality image can be quantified by this metric.}
    \label{fig:rd-lpips}
\end{figure}

All common reconstruction losses have been blamed for not being close to human perceptual vision \cite{zhao2017losses}. L1 loss pays more attention to edges and high frequency areas of the image, L2 loss results in blurry reconstruction and SSIM/MS-SSIM losses will have an effect of not reconstructing minute details like text in images.
Training our autoencoder based on MSE and LPIPS will result in  outperforming even the state of the art hand-engineered codec, i.e., BPG \cite{bpg}, as shown in Figure \ref{fig:rd-cvrves}.

The general lower performance of our GAN network is a common issue addressed in \cite{blau2018}. The PSNR or MS-SSIM are unable to capture the perceptual quality of the generated image in a GAN. The perceptual quality of GAN network reconstructions is discussed in section \ref{sec:experiments:ablation}, measured by perceptual metrics.
%Although at very low bit-rates, as generative network tries to keep the distribution of its input image at the reconstructed image, it may deviate a bit from the true pixel-wise values of the input pixels which results in degraded performance of the GAN network in comparison to the vanilla attention-based autoencoder.

\subsection{\textbf{Ablation Study}} \label{sec:experiments:ablation}
To investigate how much attention modules contribute to the performance of our neural compressor, we have trained three separate networks for each of the seven targeted bit-rates discussed in section \ref{sec:experiments:implementation}. The first architecture has only the WNLAM module (Figure \ref{fig:rd-cvrves}) whose performance in terms of PSNR and MS-SSIM has been improved by adding the WCBAM attention module.

%Impact of placing the decoder into the generator of a GAN network and scoring the outputs of the generator by the discriminator as shown in Figure \ref{fig:network-arch}. 
As emphasized in Figure \ref{fig:visual-comparison}, the adversarially trained decoder results in better visual quality on the reconstructed image than the autoencoder only trained with attention mechanism. Conventional metrics like PSNR and MS-SSIM are unable to capture the higher perceptual quality of the GAN reconstructed images. We empirically found that LPIPS can show the merit of adversarially trained network. As it is shown in Figure \ref{fig:rd-lpips}, LPIPS values correspond to the human judgment of the quality of reconstructed images.

\section{\textbf{Conclusion}}\label{sec:conclusion}
In this work, we have shown how an effective image compression scheme based on trainable neural networks could be utilized for ad-hoc applications like images from NASA's SDO mission. In addition, we explored the effectiveness of attention mechanisms in an adversarially trained neural network to improve performance of compression in terms of rate-distortion-perception trade-off.

%\section*{\textbf{Acknowledgment}}
%This work was supported by NASA EPSCoR program.
\bibliographystyle{IEEEtran}
\bibliography{mybib}
\end{document}